\documentclass[prl,aps,amssymb,showpacs,twocolumn]{revtex4}

\usepackage{dcolumn}
\usepackage{graphicx}
\usepackage{bm}

\usepackage{amsmath}
\usepackage{amssymb}
\usepackage{amsthm}
\usepackage{amsfonts}
\usepackage{enumerate}
\usepackage{latexsym}

\input{epsf}

\begin{document}
\title{Collective modes of a helical liquid}
\author{S. Raghu, Suk Bum Chung, Xiao-Liang Qi and Shou-Cheng Zhang}
\affiliation{Department of Physics, McCullough Building, Stanford
University, Stanford, CA 94305-4045}
\date{\today}
\begin{abstract}
We study low energy collective modes and transport properties of the
``helical metal" on the surface of a topological insulator.  At low
energies, electrical transport and spin dynamics at the surface are
exactly related by an operator identity equating the electric
current to the in-plane components of the spin degrees of freedom.
From this relation it follows that an undamped spin wave always
accompanies the sound mode in the helical metal -- thus it is
possible to `hear' the sound of spins. In the presence of long range
Coulomb interactions, the surface plasmon mode is also coupled to
the spin wave, giving rise to a hybridized ``spin-plasmon" mode. We
make quantitative predictions for the spin-plasmon in ${\rm
Bi}_2{\rm Se}_3$, and discuss its detection in a spin-grating
experiment.
\end{abstract}


\maketitle \emph{Introduction - } Recently, topological insulators
have been theoretically predicted and experimentally observed in
both quasi-two dimensional (2D) and three dimensional (3D)
systems\cite{bernevig2006d,koenig2007,fu2007,hsieh2008,zhang2009,xia2009,chen2009}.
The concept of a topological insulator can be defined within the
non-interacting topological band theory\cite{fu2007b,moore2007} or
more generally within the topological field theory\cite{qi2008},
which is also valid for interacting systems. The simplest
topological insulators such as ${\rm Bi}_2{\rm Se}_3$ and ${\rm
Bi}_2{\rm Te}_3$ have a full bulk insulating gap and a surface state
consisting of a single Dirac cone\cite{zhang2009,xia2009,chen2009}.
As is the case for the helical edge states of a 2D topological
insulator\cite{wu2006,xu2006}, the spin and the momentum are
intimately locked in the ``helical metal" surface state of the 3D
topological insulator. This locking effect has been theoretically
predicted\cite{zhang2009} for ${\rm Bi}_2{\rm Se}_3$ and ${\rm
Bi}_2{\rm Te}_3$, and experimentally observed\cite{Hsieh2009a} in
${\rm Bi}_2{\rm Se}_3$.

In this paper, we study the universal surface state properties of
the simplest 3D topological insulators,  and consider a system
governed by a single isotropic Dirac cone at energy scales much
lower than the bulk insulating gap.  We study the consequences of
the helical nature of the metallic states: the coupling between spin
and charge excitations, and collective modes of the helical liquid.
Our theory here is directly applicable to the case of the  ${\rm
Bi}_{2}{\rm Se}_3$/${\rm Bi}_2{\rm Te}_3$ family, which has a single
isotropic Dirac cone that remains isotropic for low dopant
concentrations.

\emph{Spin dynamics and electrical transport  - } Starting from a
low energy effective Hamiltonian for the bulk of a 3D topological
insulator in the ${\rm Bi_2Se_3}$ family, the low energy surface
Hamiltonian was derived by Zhang {\it et al. }\cite{zhang2009} by
diagonalizing the bulk effective Hamiltonian with open boundary
conditions, and by integrating out the high energy bulk degrees of
freedom. In Ref. \cite{zhang2009}, it was shown that for a surface
in the xy-plane, the helical  states at low energies are governed by
\begin{equation}
H = \int{d^2{\bf x}}\psi^\dagger({\bf x}) \left[\hbar v_f ({\bf
\hat{z}} \times (-i\nabla))\cdot{\bm \sigma}-\mu\right]\psi({\bf
x}),
\end{equation}
where $\bf k$ is the Bloch vector in the 2D surface Brillouin zone
and $\bm \sigma$ the Pauli matrices describing electron spin. Here,
$\mu$ is the value of the chemical potential relative to the surface
Dirac point. This Hamiltonian adequately captures the dynamics of
the surface for energies much smaller than the bulk gap $\Delta$,
and for length scales much larger than
$\hbar v_f/ \Delta$, the penetration depth of the gapless surface states into the bulk. 
In this regime, there are several universal features that
characterize the surface states.  First there is the operator
identity relating the charge current to  the in-plane component of
the spin on the surface:
\begin{equation}
\label{operatoridentity} {\bf j}({\bf x}) = \psi^\dagger({\bf x}) v_f \left(\bm \sigma \times {\bf \hat{z}}
\right)\psi({\bf x})\equiv v_f{\bf S}({\bf x})\times\hat{\bf z}
\end{equation}
Such a simple relation between charge current density and spin
density is the key observation of this work, which is a unique
property of the helical liquid, and leads to many intrinsic
correspondences between spin and charge dynamics in the system. It
should be noticed that the operator identity
Eq.~(\ref{operatoridentity}) remains valid even when interaction
terms are added to the Hamiltonian. The only deviation from this
identity comes from the change of Fermi surface shape due to
rotational symmetry breaking, which has been observed in ${\rm
Bi_2Te_3}$ when the chemical potential is close to the bottom of the
 bulk conduction band \cite{chen2009}.
In this regime, the Fermi velocity obtains an $O(k/k_f)^2$
distortion\cite{Fu2009, Alpichshev2009}, and the operator identity above no longer
rigorously holds. However, there is still a one-to-one
correspondence between the velocity and spin of the surface
electrons, so that many conclusions we will discuss in the following
will still hold qualitatively in this regime.

In what follows, we will assume a perfectly circular fermi surface
in the continuum limit. From the above identity
(\ref{operatoridentity}), we derive the important dynamical identity
between correlation functions
\begin{equation}
v_f^2 \epsilon_{ik} \epsilon_{jl} \langle T_{\tau} s_i( \tau) s_j (0) \rangle = \langle T_{\tau} j_k(\tau) j_l (0) \rangle
\end{equation}
which relates the electric transport to the dynamical spin structure
factor in the linear response regime. The electric response of the
helical metal can thus be calculated from the generalized spin and
density susceptibility
\begin{equation}
\chi_{ \mu \nu}({\bf q},i\Omega_m\!) \!=\!\! \frac{-1}{\beta}\!\sum_{{\bf k},i\omega_n}\!\!\! {\rm Tr}[\sigma_\mu\mathcal{G}({\bf k}+{\bf q},i\omega_n + i\Omega_m)\sigma_\nu\mathcal{G}({\bf k}, i\omega_n)],
\label{EQ:suscepTensor}
\end{equation}
where $\sigma_\mu = (1,{\bm \sigma})$,
and we have introduced the Matsubara Green function $\mathcal{G}$.
In particular, the optical conductivity $\sigma_{xx}(\omega)$ is
simply related to dynamical spin susceptibility:
$\sigma_{xx}(\omega)=\frac{i}{\omega}v_f^2\chi_{yy}(\omega)$, while
the Hall conductivity is related to the off-diagonal dynamical spin
susceptibility
$\sigma_{xy}(\omega)=-\frac{i}{\omega}v_f^2\chi_{yx}(\omega)$. In
principle, this relation can be verified by comparing the optical
conductivity with the experimental results of spin-polarized neutron
or polarized light scattering. 

Due to the rotational invariance of the Fermi surface about the
$\hat{z}$ axis, the longitudinal ($s^L ={\bf \hat{q}}\cdot{\bf s}$)
and transverse ($s^T = {\bf \hat{z}}\cdot({\bf \hat{q}}\times {\bf
s})$) components of the spin excitations  are decoupled.  In
particular, it can be shown that susceptibilities involving
correlations between longitudinal and transverse spin degrees of
freedom vanish identically. An explicit evaluation of
Eq.\eqref{EQ:suscepTensor} shows that the $4\times 4$ susceptibility
tensor decomposes into two $2\times 2$ matrices, one with the
density and transverse spin, and the other with the perpendicular
($s_z$) and the longitudinal spin.  The density-transverse spin
block of the susceptibility tensor can be determined from the Ward
identity that has its origin in the continuity equation for the density:
$\partial_t n_{\bf q} = -iq j^L_{\bf q}$, where $j^L_{\bf q} = {\bf
\hat{q}}\cdot{\bf j}_{\bf q}$ is the longitudinal current.  The Ward
identity, combined with the operator identity in
Eq.\eqref{operatoridentity} gives us
\begin{equation}
\partial_t n_{\bf q} = -iq s^T_{\bf q}.\label{continuity}
\end{equation}
We note that the longitudinal current is simply the transverse spin
degree of freedom.

The structure of the density and spin response bears a remarkable
resemblance to source-free Maxwell electrodynamics in $2+1$
dimensions, as first pointed out in Ref. \cite{bernevig2005}. The
analogy is made precise when we identify the density with the
``magnetic field" perpendicular to the surface, and the transverse
spin components with the ``electric field" in the plane of the
surface. Eq. (\ref{continuity}) can be written as $\partial_t
n=-\nabla\times {\bf s}^T$ in the real space. The equation of
continuity that connects the density to the transverse spin is thus
precisely the Faraday law for this system. Moreover, Gauss' law for
the electric field $\nabla\cdot {\bf s}^T=0$ is satisfied by
construction, since the spins are transverse to the in-plane
momentum. A similar identification has been made for a system with
Rashba spin-orbit coupling \cite{bernevig2005}; however, the analogy
to electrodynamics is less precise in Rashba systems than in the
case of the helical metal due to the presence of the quadratic
dispersion in the Rashba Hamiltonian.

Applying the Ward identity to the response functions, we find that the
density and transverse spin subset of the susceptibility tensor in the basis
$\left( n, s^T \right)$ has the
form
\begin{equation}
\label{chi}
{\bf \chi} = \left(\begin{array}{cc} 1                     & x\\
                                     -x &  -x^2 \end{array}\right)\chi_{nn}, \ \  x = \frac{\omega}{v_fq}
\end{equation}
We note that in order to obtain this form, we need to regularize the
susceptibility tensor, taking the $\mu=0$ ground state as the vacuum
\cite{Polini2009, Principi2009}.

\begin{figure}[bht]
\centering
   \includegraphics[width=.4\textwidth]{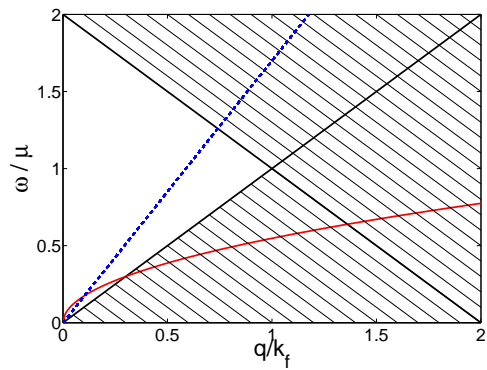}
\caption{Dispersion of the collective modes in the long wavelength regime.  The plasmon
(solid red line) is shown for $\alpha = 0.6 $ and the zero-sound (dashed blue line) mode
is shown for $U=0.3 U_c$.  The
shaded region denotes the particle-hole continuum where the imaginary parts of the
density and spin susceptibilities are non-zero.  The collective modes are long-lived in the
unshaded region  $v_f q < \omega < 2 \mu - v_f q$.  }
\label{FIG:omegaq}
\end{figure}

The bare susceptibilities can be evaluated explicitly for arbitrary
$\omega $ and $q$ using the non-interacting Green function
\begin{eqnarray}
\mathcal{G}^{(0)}({\bf k}, i\omega_n) &=& [i\omega_n - \mathcal{H}({\bf k})]^{-1} = \sum_{s=\pm 1}\mathcal{P}_s G_s^{(0)}(i\omega_n, {\bf k}), \nonumber \\
\mathcal{P}_s &=& \frac{1}{2}[1+s({\bf \hat{z}}\times{\bf \hat{k}})\cdot{\bm \sigma}],\nonumber\\
G_s^{(0)} &=& [i\omega_n-sv_f k+\mu]^{-1}
\end{eqnarray}
As shown in Fig. \ref{FIG:omegaq}, we find that
particle-hole excitations are found everywhere except for $v_f q <
\omega < 2 \mu - v_f q$. In this region, the imaginary
components of all the susceptibilities vanish and collective
excitations are undamped.

{\it Collective modes -}
In the long-wavelength, low frequency limit, $q \ll k_f, \omega \ll \mu$,
the non-interacting density susceptibility to leading order is
\begin{equation}
\chi_{nn} = \frac{\mu}{2 \pi \hbar^2 v_f^2} \left[ \frac{x}{\sqrt{x^2 - 1}} - 1 \right], \  x = \frac{\omega+i \delta}{v_f q}
\end{equation}
with corrections $O(q/k_f)^2$.  Thus, for $\omega > v_f q $,
 the imaginary part of the density and spin
susceptibilities vanish and undamped collective modes exist. To
determine the collective mode spectrum in this regime, we treat the
susceptibility tensor within the random phase approximation (RPA).

In the case of the long-ranged Coulomb interaction, $U(q)$,
the RPA correction to the susceptibilities in Eq.\eqref{chi} is given
by
\begin{equation}
\label{coulomb}
\hat{\chi}^{RPA} ({\bf q}, \omega + i \delta ) = \hat{\chi} \left[ \hat{1} - \frac{U(q)}{2} \left( \hat{1}+  \hat{\tau}^z \right)  \hat{\chi} \right]^{-1}
\end{equation}
where $\hat{\chi}$ is the density and transverse spin susceptibility
in Eq.\eqref{chi} and $\tau^z$ is a Pauli matrix in the same basis.
Although the susceptibility is a $2\times 2$ matrix, the result of
the RPA correction turns out to be the same as an ordinary 2D Fermi
liquid:
\begin{equation}
\hat{\chi}_{RPA} = \left(\begin{array}{cc} 1                     & x\\
                                     -x &  -x^2 \end{array}\right)\frac{\chi_{nn}}{1-U\chi_{nn}}, \ \  x = \frac{\omega}{v_fq}
\end{equation}
Collective mode excitation spectra are obtained via the poles of the
matrix of RPA susceptibilities. For the Coulomb interaction $U(q) =
2\pi\hbar\alpha v_f/q$, where $\alpha = e^2/\epsilon_d h v_f$
($\epsilon_d$ being the dielectric constant of the topological
insulator) is the fine structure constant for the helical metal. 
Therefore the plasmon dispersion satisfies
\begin{equation}
\frac{\hbar v_f q}{\alpha \mu} = \frac{x}{\sqrt{x^2 - 1}} - 1
\label{EQ:CoulombPole}
\end{equation}
which for small $q$ establishes that plasmon modes are gapless and
propagate with a dispersion
\begin{equation}
\omega = \frac{\mu}{\hbar}\sqrt{ \frac{\alpha}{2}\frac{q}{k_f}},
\end{equation}
similar to an ordinary Fermi liquid in 2D. Note that in order for
the plasmon to propagate, the solution to Eq.\eqref{EQ:CoulombPole}
needs to stay in the unshaded region
of Fig.~\ref{FIG:omegaq}, 
which means that for frequency above $\omega_c = (\alpha/2)(\mu/\hbar)$,
the plasmon will merge into the particle-hole continuum. We
expect $\hbar\omega_c \sim $2.2meV for the Bi$_2$Se$_3$ family;
this comes from $\epsilon_d \approx 100$ \cite{RICHTER1977} and
$v_f \approx 5.0 \times 10^5$m/s with $\mu$ typically around
100meV \cite{xia2009,Hsieh2009a}.

\begin{figure}[bht]
\centering
   \includegraphics[width=.4\textwidth]{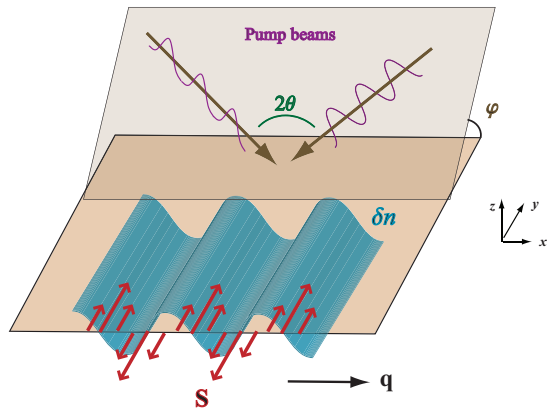}
\caption{The spin-plasmon collective mode: a density fluctuation
(green surface) induces a transverse spin fluctuation (red arrows) or {\it vice-versa}.
To detect the spin-plasmon, a spin-density wave is generated by a
spin grating. Two orthogonally polarized non-collinear incident
beams (relative angle of $2\theta$) induce a spin polarization wave.
The transverse component of the photon helicity is $P \cos \theta
\cos \varphi$ where $P$ is the total photon helicity amplitude and
$\varphi$ is the beam plane makes with the surface. The induced
plasmon charge oscillation can be detected by conventional means.}
\label{FIG:grating}
\end{figure}

In sharp contrast to the conventional Fermi liquid, the surface
plasmon of the helical liquid always carries spin, and could be
appropriately called the {\it spin-plasmon}.  This is a consequence
of the equation of continuity for the density and the operator
identity, Eq. \ref{operatoridentity}: $\partial_t n_{\bf q} = -iq
s^T_{\bf q}$. Thus, a fluctuation in the density will be accompanied
by a transverse spin fluctuation due to the ``Faraday law" that
couples them. In Fig. \ref{FIG:grating}, the nature of the
spin-plasmon is shown pictorially at a fixed $\omega$ and $q$.  A
density oscillation induces a transverse spin wave in perfect
analogy with Maxwell electrodynamics in $2+1$ dimensions. At the
peaks and troughs of the density wave, the transverse spin
components are zero, whereas regions where the density variations
vanish are accompanied by a maximum spin polarization.  The locking
of the spin and charge degrees of freedom in the spin-plasmon
collective mode is unique to the helical liquid and marks a striking
difference relative to the ordinary Fermi liquid.

For short-ranged Hubbard-like interactions, the RPA susceptibilities
are
\begin{equation}
\label{shortrange}
\hat{\chi}^{RPA} ({\bf q}, \omega + i \delta ) = \hat{\chi} \left[ \hat{1} - U \hat{\tau}^z \hat{\chi} \right]^{-1}
\end{equation}
The reason for this difference in the RPA susceptibilities is that
short ranged Hubbard-like interactions can be decomposed in both the
density and spin channels, both of which contribute to RPA diagrams,
whereas the Coulomb interaction can only be decomposed in the charge
channel in the RPA approximation. In both cases, the RPA
susceptibilities satisfy the Ward identity.

The zero sound spectra for short range interaction can be determined
by the pole of Eq. (\ref{shortrange}), which is given by
\begin{equation}
\frac{2 \pi \hbar^2 v_f^2}{U \mu} = \left( x^2 + 1 \right) \left(
\frac{x}{\sqrt{x^2 -1}} - 1 \right)\label{soundmodeeqn}
\end{equation}
In this case,
there is an essential difference from the ordinary fermi liquid. The
right hand side of Eq. (\ref{soundmodeeqn}) takes its value in the
range $[\frac12,+\infty)$, so that the equation has a solution only if
${4 \pi \hbar^2 v_f^2}/{U \mu} > 1$, or $U<U_c=4 \pi \hbar^2 v_f^2 /
\mu$. The sound wave dispersion for the two limiting cases $U\ll
U_c$ and $U\lesssim U_c$ are
\begin{equation}
\omega = \left\{\begin{array}{cc}
v_fq\left(1+\frac{8U^2}{U_c^2}\right),&
U/U_c\rightarrow 0\\v_f q \sqrt{\frac{7}{4}\frac{U}{U_c-U}}
,&U/U_c\rightarrow 1\end{array}\right.
\end{equation}
Physically, the disappearance of the sound mode for strong interaction
$U>U_c$ can be understood as a consequence of the attractive
interaction in the spin channel: note that the interaction vertex in Eq. (\ref{shortrange})
for  the density correlation  $\chi_{nn}$ has the opposite sign of the vertex
for the transverse spin correlation $\chi_{TT}$.
In an ordinary Fermi liquid with repulsive short range interaction,
the same argument leads to the damping of the  spin wave, but does not
affect the zero sound since the spin and charge response are
decoupled (or only weakly coupled in systems with weak to moderate spin-orbit coupling).
By contrast, in the helical spin liquid, the spin and
charge responses are intrinsically coupled due to the operator
identity Eq.~(\ref{operatoridentity}), so that the sound wave can be
damped even for a repulsive interaction.





\emph{Spin grating measurements - }
Our basic method of detecting the spin-plasmon mode is to
excite the spin degree of freedom and to detect the propagating density
wave coupled to the spin polarization wave. In other words, we
generate transverse spin polarization and detect the induced density
wave, which we can measure through spatial modulation of
reflectivity.

For generating the transverse spin wave, we propose a method similar
to the transient spin grating (TSG) used in
Refs. \onlinecite{WEBER2007} and \onlinecite{KORALEK2009}. For our
version of TSG, shown in Fig.~\ref{FIG:grating}, we need to have two
non-collinear femtosecond laser beam pulses incident on the
topological insulator surface, with $\hat{\bf z}$ set as the surface
normal; the wave vector directions of these two beams are
$(\pm\sin\theta,-\cos\theta\cos\varphi,-\cos\theta\sin\varphi)$. By
linearly polarizing the two beams in orthogonal directions, we can
generate by interference alternating photon helicity in the
direction $(\sin \theta, -\cos\theta \cos \varphi, -\cos \theta
\sin\varphi)$ with the grating vector ${\bf q}$ along $\hat{\bf x}$;
the $q$ can be varied by changing the relative phase between the two
interfering beams. For the persistent spin helix experiment,
Ti:sapphire lasers (wavelength 650$\sim$1100nm) were used to obtain
$q=0.34-2.5\times10^4$cm$^{-1}$ \cite{KORALEK2009}; this $q$ value
would suit our purpose (a typical $k_f$ with $\mu$ lying in the bulk
gap can be taken as $60\times10^4$cm$^{-1}$ for ${\rm Bi}_{2}{\rm
Se}_3$ or ${\rm Bi}_2{\rm Te}_3$ \cite{xia2009,chen2009}). So long
as we keep the two beams' polarization orthogonal and intensity
equal, this photon helicity wave does not directly generate any
density wave. However, the spin polarization wave it generates will
have a nonzero $s^T\propto \cos\theta\cos\varphi$ for
$\varphi \neq \pi/2$.  
After the laser pulses are applied, the transverse spin components
$s^T$ propagate through the spin-plasmon collective mode. On the
other hand, other spin components are Landau damped. Note that
intensity $I$ of the induced density-spin wave has $\varphi$
dependence, which vanishes for $\varphi=\pi/2$ since in that case
$s^T = 0$. Observation of such dependence would be a strong evidence
confirming our helical liquid theory.

In conclusion we presented a general theory of the collective modes
of the helical liquid. We derived a general relation between the
charge current and the spin which is valid for arbitrarily
interacting systems. We show that the the spin-plasmon mode
propagates on the surface of a topological insulator, and propose a
experimental setting to detect this mode. The spin-plasmon mode
unifies spintronics and plasmonics, two frontier branches of current
research, and can be used for spin transport in spintronics devices.

We are grateful to Mark Brongersma, Kai Chang, Run-Dong Li and Joseph Orenstein for insightful discussions.
This work is supported by the Department of Energy, Office of Basic
Energy Sciences, Division of Materials Sciences and Engineering,
under contract DE-AC02-76SF00515, and by the Stanford Institute
for Theoretical Physics (SITP) postdoctoral fellowship (SR and SBC).
Part of this work was carried out at the Kavli Institute for Theoretical
Physics, UC Santa Barbara, with support from KITP's NSF
Grant No.   PHY05-51164.

\bibliography{collective}

\end{document}